\documentclass[twocolumn,twoside,slac_two]{revtex4}
\usepackage{graphicx}
\usepackage{fancyhdr}
\def\Journal#1#2#3#4{{#1} {\bf #2}, #3 (#4)}

\def\NIMA{{Nucl.\ Instrum.\ Methods} A}

\def\PLB{{Phys.\ Lett.\ }  B}
\def\PRL{Phys.\ Rev.\ Lett.\ }
\def\PRD{{Phys.\ Rev.\ } D}

\def\be{\begin{equation}}
\def\ee{\end{equation}}
\def\bea{\begin{eqnarray}}
\def\eea{\end{eqnarray}}

\def\epem{e^+e^-}

\def\BBbar{B\bar{B}}

\def\KS{K^0_S}

\def\GeV{{\rm~GeV}}

\def\Mbc{M_{\rm bc}}

\def\DeltaE{\Delta E}

\def\bsig{B_{\rm sig}}
\def\btag{B_{\rm tag}}

\def\PM#1#2{\,^{+#1}_{-#2}}

\def\etal{\textit{et al.}}

\def\g{\gamma}
\def\l{\ell}

\def\blnu{B^+ \rightarrow  \l^+ \, \nu_\l}
\def\btaunu{B^+ \rightarrow \tau^+ \, \nu_\tau}
\def\bmunu{B^+ \rightarrow \mu^+ \, \nu_\mu}
\def\benu{B^+ \rightarrow e^+ \, \nu_e}
\def\blnug{B^+ \rightarrow  \l^+ \, \nu_\l \, \g}

\def\be{\begin{equation}}
\def\ee{\end{equation}}
\def\ba{\begin{eqnarray}}
\def\ea{\end{eqnarray}}
\def\nub        {\ensuremath{\overline{\nu}}}
\def\tautoenunu {\ensuremath {\tau^+ \to e^+ \nu \nub}}
\def\tautomununu {\ensuremath {\tau^+ \to \mu^+ \nu \nub}}
\def\tautopinu {\ensuremath {\tau^+ \to \pi^+ \nub}}
\def\tautopipiznu {\ensuremath {\tau^+ \to \pi^+ \pi^{0} \nub}}
\def\btodlnux {\ensuremath{B^- \to D^0 \ell^{-} \bar{\nu}_{\ell} X}}
\def\btodstlnu {\ensuremath{B^- \to D^{\ast 0} \ell^{-} \bar{\nu}_{\ell} }}
\def\eextra {\ensuremath{E_{\mathrm{extra}}}}
\begin{document}

\title{\boldmath Review of $B_u$ leptonic decays}


\author{Stefano Villa}
\affiliation{Laboratory for High Energy Physics,
Swiss Federal Institute of Technology, 1015 Lausanne, 
Switzerland}

\begin{abstract}
This paper reviews the status of searches and measurements
of $B_u$ leptonic decays, concentrating on the most recent
results obtained at $B$--factories. We will describe studies of decays
of the type $\blnu$ and $\blnug$.
\end{abstract}

\maketitle

\thispagestyle{fancy}

\section{Introduction}
%%%%%%%%%%%%%%%%%%%%%%%%%%%%%%%%%%%%%%%%%%%%%%%%%%%%%%%%%%%%%%%%%%%%%%%%%%%
%
There are several reasons for studying purely leptonic decays of charged
$B$ mesons. 
Such processes are rare but they have clear experimental signatures due
to the presence of a highly energetic lepton in the final state. 
The theoretical predictions are very clean due
to the absence of hadrons in the final state. These features make such
decays the ideal ground to look for deviations from the Standard Model (SM) 
predictions hoping for some signs of New Physics (NP). 
Furthermore, as described in Sec.~\ref{Sec:blnu}, measuring the
branching fraction (BF) of these modes allows direct insight into some 
of the fundamental parameters of the theory, that are not easily accessible
otherwise.

In this review we will
describe some of the latest results on leptonic and radiative leptonic 
decays of charged $B$ mesons obtained at the $B$-factory experiments, 
Belle~\cite{belle} and BaBar~\cite{babar}, both collecting $\BBbar$ pairs
at the $\epem \to \Upsilon(4S)$ resonance.
The common characteristic of all the analyses described is the presence
of one (for electronic and muonic modes) or more (for the tauonic mode) neutrinos
in the signal $B$ ($\bsig$). To suppress background and constrain kinematically
the $\bsig$ reconstruction, the common solution is to reconstruct the
other $B$ meson in the event, usually referred to as tag, companion or recoil
$B$ ($\btag$ in our notation).
The $\btag$ reconstruction can be performed inclusively, by reconstructing
and identifying all particles in the event that do not belong to the
signal and using them to form a $B$ candidate, or exclusively, by selecting several 
decay modes adding up to a reasonable BF and
reconstructing $\btag$ in these modes.
A great reduction of the continuum background is obtained by selecting
events in which the $\btag$ candidate is kinematically consistent with 
a $B$ meson issued from the decay of a $\Upsilon(4S)$. The
selection is usually performed on the two following variables:
the beam constrained mass (called $\Mbc$ by Belle and $m_{\mathrm{ES}}$ by BaBar)
$\Mbc  = \sqrt{E_\mathrm{beam}^2 - |{\vec p_{B}^{\,*}}|^2}$
and $\Delta E = E_B^* -E_\mathrm{beam}$, where 
$\vec p_{B}^{\,*}$ and $E_B^*$ are the momentum and energy
of $\btag$, all variables being evaluated in the Center-of-Mass 
(CM) frame. 
For $\BBbar$ events, $\Mbc$ is centered at the value of the $B$ mass and
$\Delta E$ at zero.
The advantage of exclusive $\btag$ selections is the very powerful suppression 
of background, at the price of a relatively small signal efficiency. Inclusive
selections can give higher efficiencies, but are better suited for channels in 
which the signal signature allows a good separation from background. 
As we will describe in the following, both techniques have been pursued with
success by Belle and BaBar. 

%%%%%%%%%%%%%%%%%%%%%%%%%%%
% searches for  B-> ell nu
%%%%%%%%%%%%%%%%%%%%%%%%%%%
%
\section{\boldmath Searches for $\blnu$ decays }\label{Sec:blnu}
Purely leptonic decays of charged $B$ mesons proceed in the SM via the 
$W$-mediated annihilation tree diagram, with a branching fraction given by:
\bea \label{eq:B2lnu}
\lefteqn{ \mathrm{BF}(B^+\rightarrow \ell^+\nu_\ell) = } \nonumber  \\
& & \frac{G_F^2 m_B}{8\pi} m_\ell^2 \left(1-\frac{m_\ell^2}{m_B^2}\right)^2
f_B^2 |V_{ub}|^2 \tau_B , 
\eea
where $\tau_B$ is the $B$ meson lifetime, $f_B$
is the $B$ decay constant and $V_{ub}$ an element of the CKM-matrix.
These modes are very interesting because they 
give direct access to the product  $f_B \times V_{ub}$, from which
one can extract a measurement of $f_B$, since $V_{ub}$ is measured
in other $B$ decay modes~\cite{HFAG}. 
The SM expectation for $B^+\rightarrow \tau^+\nu_\tau$ is 
$\mathrm{BF}(B^{+}\rightarrow\tau^{+}\nu_{\tau}) 
= (1.59 \pm 0.40) \times 10^{-4}$, assuming for $V_{ub}$ the value
$(4.39 \pm 0.33) \times 10^{-3}$ determined by inclusive
charmless semileptonic $B$ decay data~\cite{HFAG}, 
$\tau_{B} = 1.643\pm 0.010$ ps~\cite{HFAG},
and $f_B = 0.216\pm 0.022$ $\GeV$
obtained from lattice QCD calculations~\cite{Gray:2005ad}.
Decays to lighter leptons are helicity suppressed, and
are their BF are predicted to be $(4.7\pm 0.7) \times 10^{-7}$
for $B^+\to\mu^+\nu_\mu$ and $(1.1 \pm 0.2) \times 10^{-11}$ for $B^+\to e^+\nu_e$.
Allowing for NP decay amplitudes, measurements of
these processes can give stringent limits on important parameters of such
SM extensions, such as the mass of the charged Higgs boson and 
$\tan\beta$ (the ratio of vacuum expectation values of the two Higgs 
doublets) in the minimal supersymmetric SM (MSSM)\cite{Hou:1992sy}.

%%%%%%%%%%%%%%%%%%%%%%%%%%%
% B-> tau nu
%%%%%%%%%%%%%%%%%%%%%%%%%%%
%
\subsection{\boldmath Searches for $\btaunu$ decay}
Both Belle and BaBar have recently presented results of searches
for the $\btaunu$ decay. Compared to the lighter leptonic modes, 
this channel has the disadvantage of additional neutrinos (one or more, 
depending on the $\tau$ decay mode).

%%%%%%%%%%%%%%%%%%%%%%%%%%%
% Belle's B-> tau nu
%%%%%%%%%%%%%%%%%%%%%%%%%%%
%
%
\begin{figure*}[t!]
\centering
\includegraphics[width=81mm]{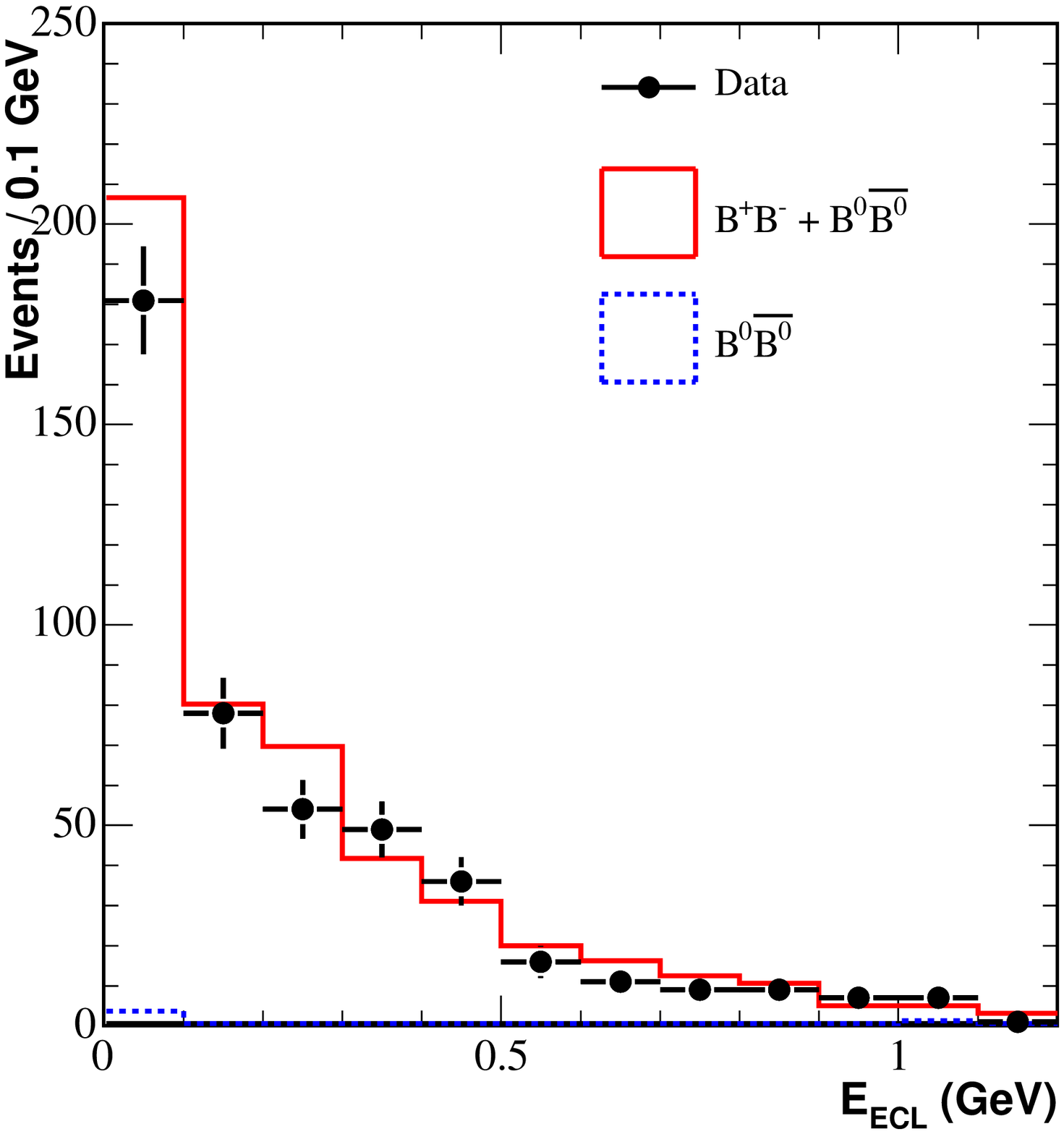}
\includegraphics[width=74mm]{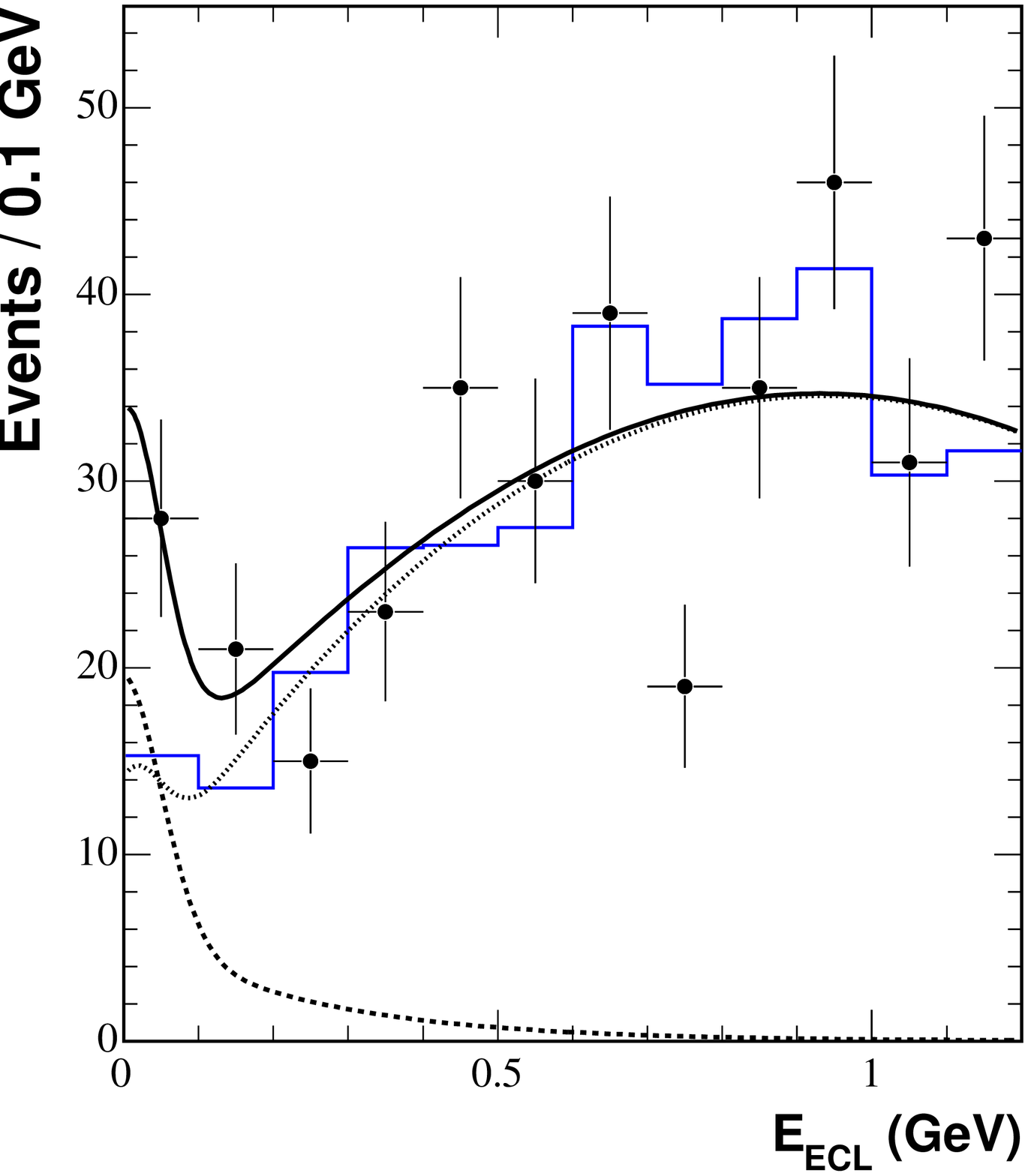}
\caption{Search for $\btaunu$ in Belle. Left: the $E_\mathrm{ECL}$ 
distribution in the double tagged control channel, showing good agreement
of MC expectation (histograms) and data (points). Right: 
$E_{\rm ECL}$ distribution in the data after
all selection criteria. 
The data and background MC samples are represented by the points 
and the solid histogram, while
the solid curve shows the result of the fit with the sum of the signal (dashed) and 
background (dotted) contributions.} \label{fig:belle-b2taunu}
\end{figure*}
\subsubsection{\boldmath Belle's first evidence for $\btaunu$ decay}
The Belle analysis~\cite{belle-btaunu} is based on a data sample of 414~fb$^{-1}$
and it employs a full reconstruction of $\btag$ in the following decay modes:
$B^{-} \rightarrow {D}{}^{(*)0} \pi^{-}$, 
${D}{}^{(*)0}\rho^{-}$, 
${D}{}^{(*)0}a_{1}^{-}$ 
and ${D}{}^{(*)0}D_{s}^{(*)-}$.
The ${D}{}^{0}$ mesons are reconstructed as 
${D}{}^{0}\rightarrow K^{-}\pi^{+}$, $K^{-}\pi^{+}\pi^{0}$,
$K^{-}\pi^{-}\pi^{+}\pi^{+}$, $K_{S}^{0}\pi^{0}$, $K_{S}^{0}\pi^{-}\pi^{+}$,
$K_{S}^{0}\pi^{-}\pi^{+}\pi^{0}$ and $K^{-}K^{+}$, and
the $D_{s}^{-}$ mesons are reconstructed as 
$D_{s}^{-}\rightarrow K_{S}^{0}K^{-}$ and $K^{+}K^{-}\pi^{-}$.
The ${D}{}^{*0}$ and $D_{s}^{*-}$ mesons are reconstructed in
${D}{}^{*0} \to {D}{}^0 \pi^0, {D}{}^0 \gamma$,
and $D_{s}^{*-} \to D_{s}^{-} \gamma$ modes. 
The $\btag$ selection yields a sample of about $6.8 \times 10^5$ $\BBbar$ 
events with a purity of 55\%.
In this sample, particles that are not assigned to $\btag$ are assigned to
$\bsig$ and a decay to a $\tau$ and a neutrino is looked for.
The $\tau$ lepton is identified in five decay modes,
$\mu^{-}\bar{\nu}_{\mu}\nu_{\tau}$,
$e^{-}\bar{\nu}_{e}\nu_{\tau}$, 
$\pi^{-}\nu_{\tau}$,
$\pi^{-}\pi^{0}\nu_{\tau}$ and 
$\pi^{-}\pi^{+}\pi^{-}\nu_{\tau}$,
which taken together correspond to $81\%$ of all $\tau$ decays. Further
background suppression is obtained by applying requirements on the
magnitude and direction of the missing momentum.

\begin{figure}[ht!]
\centering
\includegraphics[width=80mm]{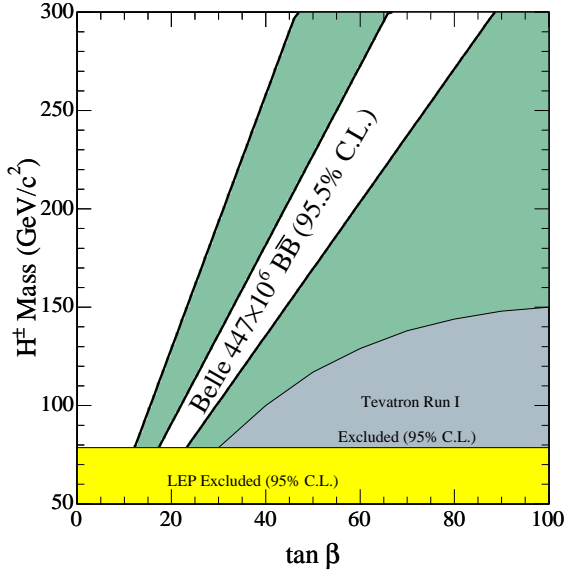}
\caption{Exclusion plot 
at 95\% C.L. in the $m_H^\pm$-$\tan\beta$ 
plane derived from the Belle measurement of the $\btaunu$ BF. The region excluded
by Belle is in green. Areas excluded by other experiments are also shown in different
colors.
} 
\label{fig:belle-mhtanb}
\end{figure}

The most powerful variable for separating signal and background is the 
remaining energy in the electromagnetic calorimeter (ECL), 
denoted as $E_{\rm ECL}$, which is the sum of
the energies of neutral clusters that are not associated with either the 
$\btag$ or the $\pi^{0}$ candidate from the 
$\tau^{-}\rightarrow \pi^{-}\pi^{0}\nu_{\tau}$ decay.
For signal events, $E_{\rm ECL}$ must be either zero or a small value 
arising from beam background hits, therefore, signal events peak at 
low $E_{\rm ECL}$.
On the other hand, background events are distributed toward higher 
$E_{\rm ECL}$ due to the contribution from additional neutral clusters.
A validation of the $E_{\rm ECL}$ simulation is performed using a control
sample of double tagged events where the $B_{\rm tag}$ is fully reconstructed 
as described above and $B_{\rm sig}$ is reconstructed in the decay chain, 
$B^{+} \rightarrow \overline{D}{}^{*0}\ell^{+}{\nu}$ 
($\overline{D}{}^{*0}\rightarrow \overline{D}^{0}\pi^{0}$),
followed by $\overline{D}^0 \to K^+ \pi^-$ or $K^+ \pi^- \pi^+ \pi^-$
where $\ell$ is a muon or an electron.
Fig.~\ref{fig:belle-b2taunu} (left) shows the $E_{\rm ECL}$ distribution in the
control sample for data and the scaled Monte Carlo (MC) simulation, 
with very good agreement between the two.
Fig.~\ref{fig:belle-b2taunu} (right) shows the $E_{\rm ECL}$ distribution obtained 
after all selections are applied and with all $\tau$ decay modes combined.
One can see a significant excess of events in the $E_{\rm ECL}$ signal region
below $E_{\rm ECL}< 0.25$ GeV. The signal yield is extracted by fitting 
the $E_{\rm ECL}$ distributions to the sum of the expected signal 
and background shapes extracted from the MC simulations, and including
a background component peaking at $E_{\rm ECL} = 0$. 
The combined fit for all five $\tau$ decay modes gives 
$17.2^{+5.3}_{-4.7}$ signal events in the signal 
region; the corresponding BF, including systematic uncertainties is:
\be
{\rm BF}(\btaunu) 
= (1.79^{+0.56}_{-0.49}(\mbox{stat})^{+0.46}_{-0.51}(\mbox{syst}))\times 10^{-4}.
\label{eq:belletaunu}
\ee
The significance is $3.5\sigma$, representing the first evidence of the purely leptonic decay
$\btaunu$. 
Using the value of $|V_{ub}|$ from~\cite{HFAG}, Belle obtains 
$f_{B} = 0.229^{+0.036}_{-0.031}(\mbox{stat})^{+0.034}_{-0.037}(\mbox{syst})$ GeV,
the first direct determination of the $B$ meson decay constant.
A measurement of ${\rm BF}(\btaunu)$ can directly
be translated into a constraint on parameters of two-Higgs-doublet models 
of type II~\cite{Hou:1992sy}, via the formula: 
\bea
{\rm BF}_{\rm MSSM}(\btaunu) & = & 
{\rm BF}_{\rm SM}(\btaunu) \nonumber \\
 & &\times \left(1-\frac{m_B^2}{m_H^2} \tan^2\beta \right)^2
\eea
which relates the SM value to the MSSM one via a factor depending only on the
mass of the $B$ meson and of the charged Higgs ($m_H$) and on $\tan\beta$.
The Belle measurement translates in the constraints in the $m_H$--$\tan\beta$ 
plane that are illustrated in Fig.~\ref{fig:belle-mhtanb}.
%

%%%%%%%%%%%%%%%%%%%%%%%%%%%
% Babar's B-> tau nu
%%%%%%%%%%%%%%%%%%%%%%%%%%%
%
\subsubsection{\boldmath BaBar's  $\btaunu$ analyses}
The BaBar collaboration has recently presented~\cite{gritsan} 
the results of two searches
of the $\btaunu$ decay, both based on the same data sample of $383 \times
10^6$ $\BBbar$ pairs, exploiting different selections for the $\btag$ 
candidates and thus yielding statistically independent results. 

\begin{figure*}[t]
\centering
\includegraphics[width=80mm]{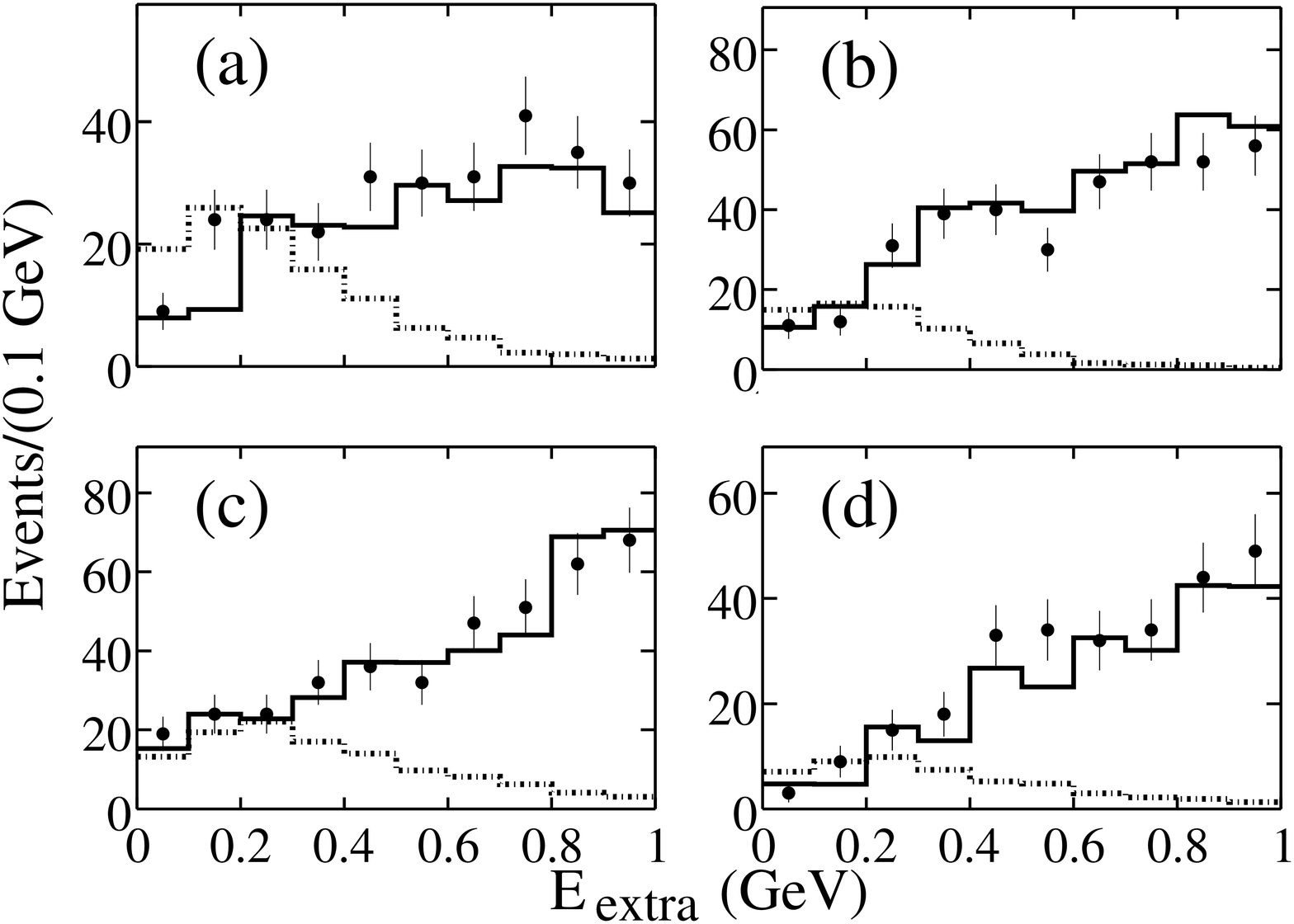}
\includegraphics[width=80mm]{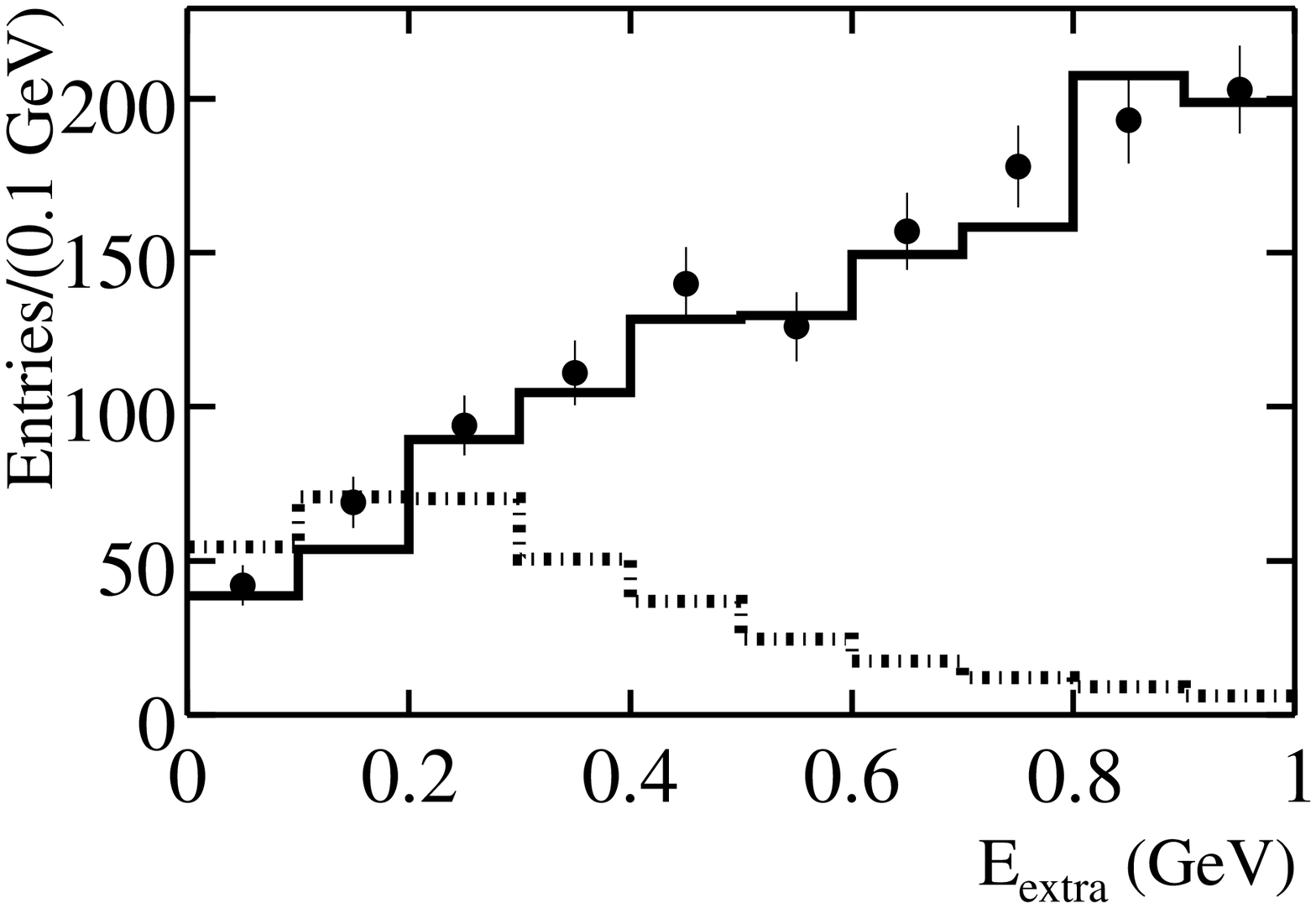}
\caption{Left: $E_\mathrm{extra}$ distributions in the BaBar semileptonig-tag 
$\btaunu$ analysis
for (a) $\tautoenunu$, (b) $\tautomununu$, (c) $\tautopinu$, and
(d) $\tautopipiznu$. Right: all four modes combined. 
Background MC (solid histogram) is rescaled to match 
the normalization of the on-resonance data (black dots) 
in the $\eextra$ sideband region. $\btaunu$ signal MC (dotted histogram) is 
normalized to a branching
fraction of $10^{-3}$ and shown for comparison.} \label{fig:babar-semilep}
\end{figure*}

The first analysis~\cite{babar-semilep} is based on the exclusive
reconstruction of semileptonic $\btag$ decays of the type 
$\btodlnux$, where $\ell$ denotes either an electron or a muon, and $X$ can be 
either nothing or a $\pi^0$ or photon from a higher mass charm 
state decay which is not explicitly included in $\btag$. 
$D^0$ candidates are reconstructed in four decay modes:
$K^{-}\pi^{+}$, $K^{-}\pi^{+}\pi^{-}\pi^{+}$, $K^{-}\pi^{+}\pi^{0}$, and
$\KS\pi^{+}\pi^{-}$. This choice of $\btag$ yields a higher efficiency 
than full reconstruction of $\btodstlnu$, but a slightly lower purity.
After selection of tagged events, the $\tau$ from the signal side is looked
for in the remaining particles of each event, in the modes 
$\tautoenunu$, $\tautomununu$, $\tautopinu$, and $\tautopipiznu$,  
constituting approximately 71\% of the total $\tau$ decay width.
Background rejection and validation of the $\btag$ selection are performed
using techniques similar to the ones described above.
The final discriminating
variable is the remaining energy ($\eextra$),
calculated by summing the CM energy of neutral clusters 
and tracks that are not associated with either $\btag$ or $\bsig$.
The $\eextra$ distributions in data and MC are shown in Fig.~\ref{fig:babar-semilep}
for the four $\tau$ decay modes separately (left) and for the combined
sample (right).
Mode dependent signal regions are chosen to optimize signal significance
in the range $\eextra < 0.25$--$0.48 \GeV$, and the expected background in 
these regions is evaluated by comparing data and MC in the $\eextra > 0.5 \GeV$
sideband and extrapolating using the MC shape.
A summary of expected background and observed signal events is presented in 
Table~\ref{tab:babar-semil}. 
\begin{table}[hbt]
\centering
\caption{\label{tab:babar-semil} BaBar's semileptonic $\btaunu$ analysis:
observed number of on-resonance data events in the signal region and 
number of expected background events.}
\begin{tabular}{lcc} \hline \hline
$\tau$            & Expected background  &  Observed events  \\ 
decay mode        & events               &  in on-resonance data  \\ \hline
$\tautoenunu$     & 44.3  $\pm$ 5.2   & 59  \\ 
$\tautomununu$    & 39.8  $\pm$ 4.4   & 43  \\ 
$\tautopinu$      & 120.3 $\pm$ 10.2  & 125  \\ 
$\tautopipiznu$   & 17.3  $\pm$ 3.3   & 18  \\ 
\hline
All modes    & 221.7 $\pm$ 12.7  & 245  \\ \hline \hline
\end{tabular}
\end{table}
Given that the signal observed is not significant, 
BaBar quotes for the BF both a measurement, 
\be
\mathrm{BF}(\btaunu) = (0.9\pm{0.6}(\mbox{stat.}) \pm 0.1 (\mbox{syst.})) \times 10^{-4}
\label{eq:semil}
\ee
and a 90\% Confidence Level (CL) upper limit,
\begin{equation}
\mathrm{BF}(\btaunu) < 1.7 \times 10^{-4}. 
\end{equation}
%

\begin{figure}[ht!]
\centering
\includegraphics[width=80mm]{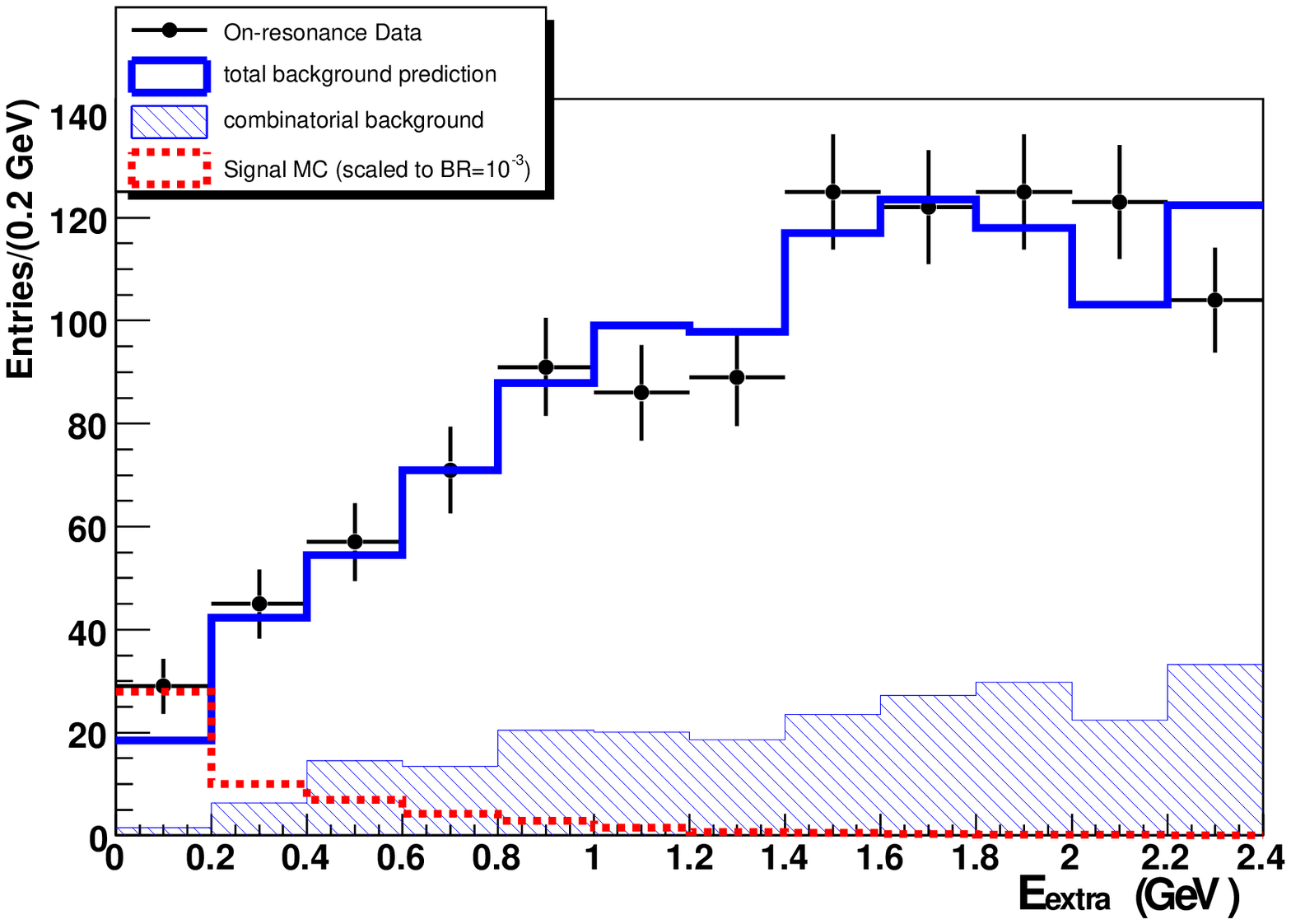}
\caption{$E_\mathrm{extra}$ distribution for the BaBar hadronic-tag 
$\btaunu$ analysis, for all $\tau$ decay modes combined.} 
\label{fig:babar-hadr}
\end{figure}

%%%%%%%%%%%%%%%%%%%%%
% Babar's hadronic tag
%%%%%%%%%%%%%%%%%%%%%
Another search for the $\btaunu$ decay is performed by BaBar by following
a strategy very similar to the one just presented, but with a different
selection of $\btag$. 
Preliminary results of this analysis were presented for the first time at 
this conference~\cite{gritsan}. 
The $\btag$ candidate is reconstructed 
in hadronic modes of the type $B^- \to D^{(\ast)0} X^-$, with 
$D^{\ast 0} \to D^0 \pi^0, D^0 \gamma$ and $X^-$ can be made of a combination of
up to five charged pions and kaons, up to two $\pi^0$ mesons and up to two
$K_S^0$ mesons. The same $\tau$ decay modes as in the semileptonic-tag study
are reconstructed in the signal side, and the results obtained are summarized
in Fig.~\ref{fig:babar-hadr}.
The slight excess of signal visible in the first bin of the plot corresponds to 
a total of 24 observed events in the (mode dependent) signal window with 
an expected background of about 14 events.
The corresponding measured BF is:
\bea
\lefteqn{\mathrm{BF}(\btaunu)  = } \nonumber  \\
 & & (1.8\PM{1.0}{0.9}(\mbox{stat.+bkg.}) \pm 0.3 (\mbox{syst.})) \times 10^{-4},
\eea
where the first error includes the statistical error and the
uncertainty on the expected background.
The significance of this measurement, including all uncertainties, is of 2.2$\sigma$.
A combination of this result with the one obtained from the semileptonic-tag analysis
(Eq.~\ref{eq:semil}) yields:
\bea
\lefteqn{\mathrm{BF}(\btaunu)  = } \\
 & & (1.20\PM{0.40}{0.38}(\mbox{stat.}) 
 \PM{0.29}{0.30}(\mbox{bkg.}) \pm 0.22 (\mbox{syst.})) \times 10^{-4} \nonumber ,
\eea
corresponding to a significance of 2.6$\sigma$.
This result is in  good agreement with the Belle measurement 
described above (Eq.~\ref{eq:belletaunu}).

%%%%%%%%%%%%%%%%%%%%%%%%%%%%%%%%%%%%%%%%%%%%%%%%%%%%%%%%%%%%%%%%%%%%%%%%%%%%%%%
%          B-> ell nu (ell = mu, e)
%%%%%%%%%%%%%%%%%%%%%%%%%%%%%%%%%%%%%%%%%%%%%%%%%%%%%%%%%%%%%%%%%%%%%%%%%%%%%%%
\subsection{\boldmath Searches for $\benu$ and $\bmunu$ decays}
\begin{figure*}[t!]
\centering
\includegraphics[width=92mm]{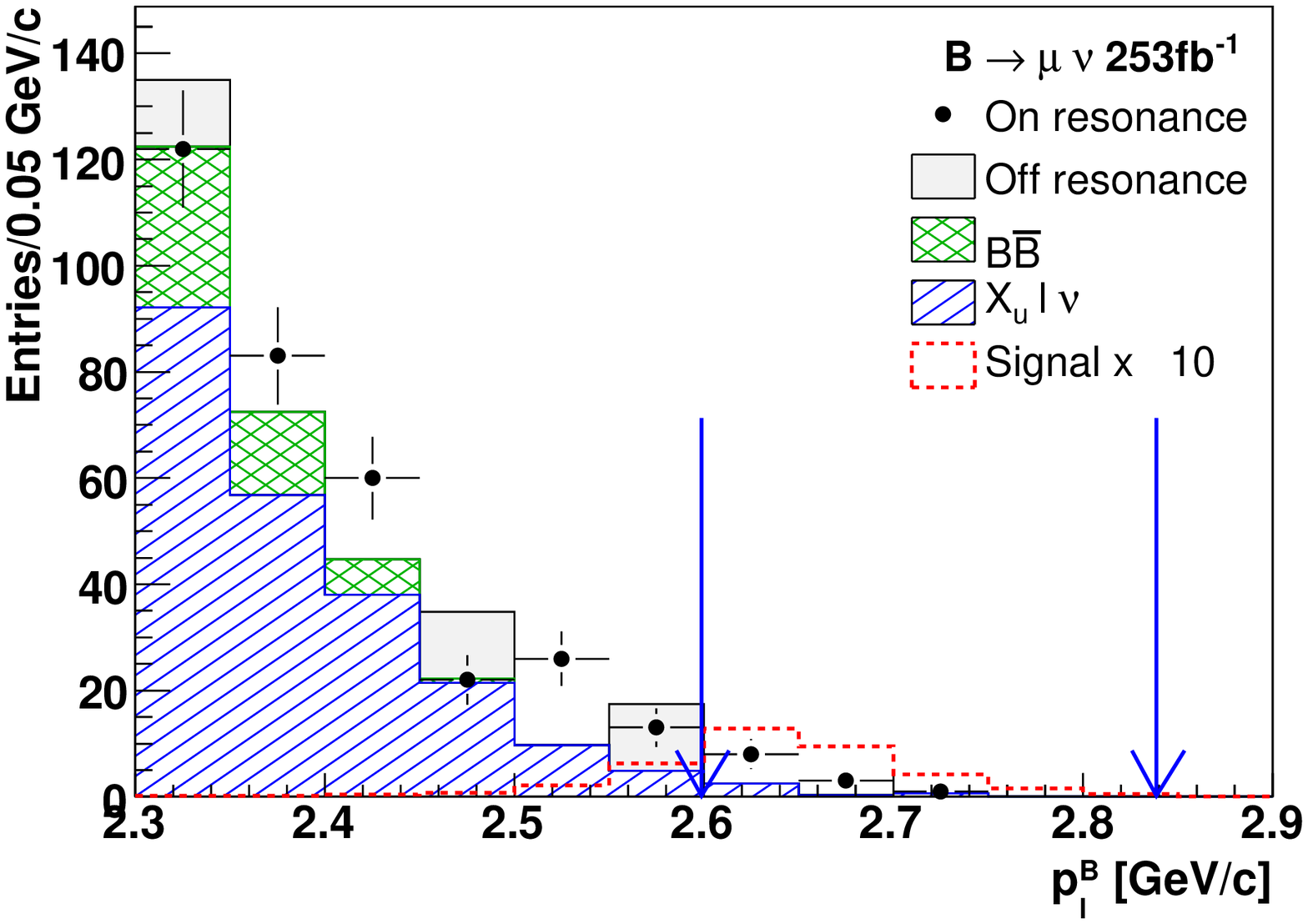}
\includegraphics[width=72mm]{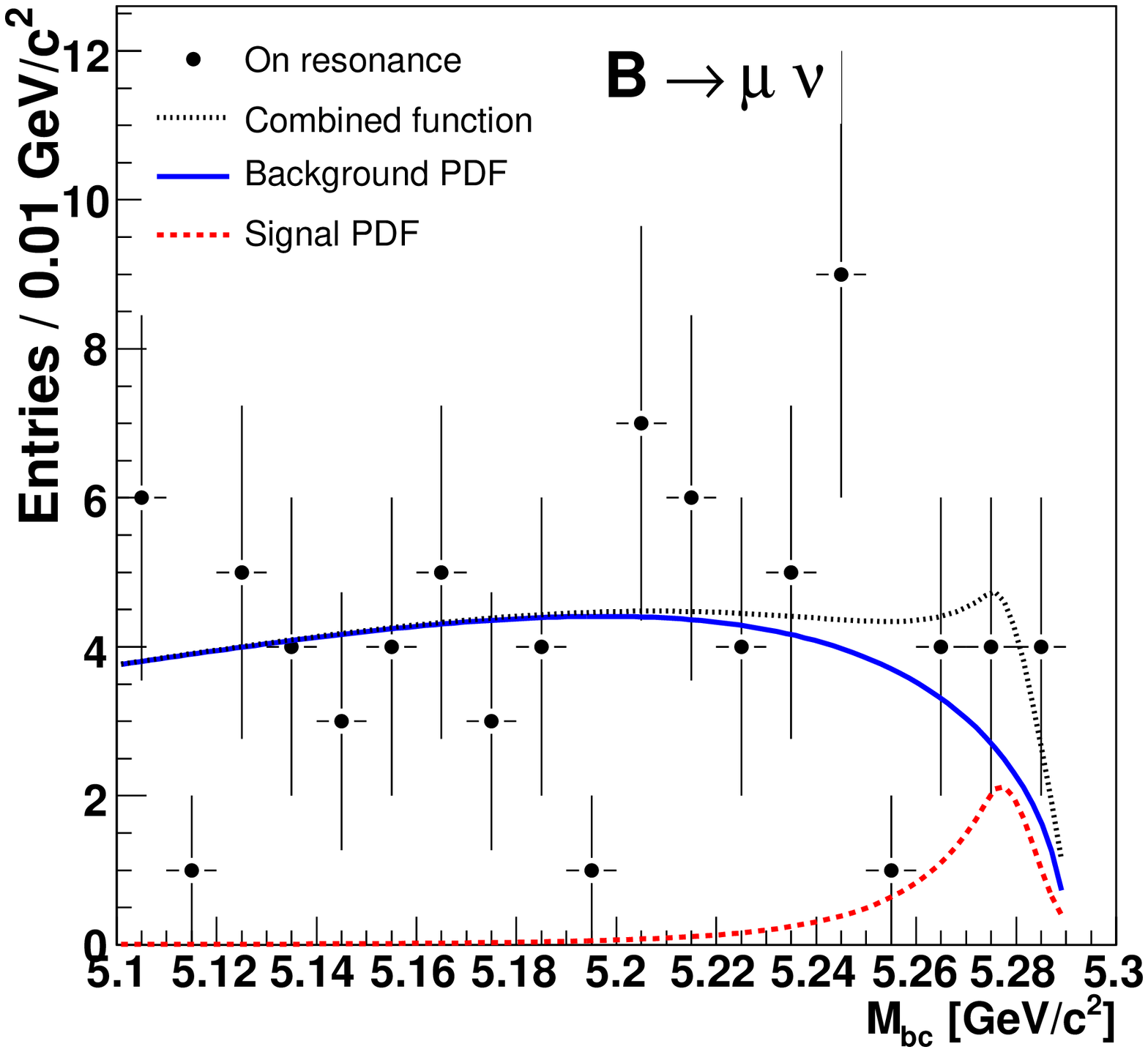}
\caption{Belle $\blnu$ analysis. 
Left: $p_\ell^B$ distributions for the signal candidates in the muon
mode.  Points show the on-resonance data, and 
solid histograms show the expected background due to
rare $B\to X_{u}\ell\nu$ decays (hatched, from MC); other 
$B\bar B$ events, principally $B\to X_c \ell \nu$ decays (cross-hatched,
also from MC); and continuum events (light shaded, taken from scaled
off-resonance data).
Dashed histograms are MC $B \to \ell \nu$ signals 
that are obtained by multiplying the SM expectations by a factor of 
10. The arrows show the signal region.
Right: $\Mbc$ distribution in the muon mode for selected events and fit result 
(dotted line) as a sum of signal (dashed) and background (solid) contributions.} 
\label{fig:belle-b2lnu}
\end{figure*}

The Belle search for $\benu$ and $\bmunu$ decays~\cite{belle-b2lnu} 
is based on an inclusive reconstruction of $\btag$, in a data sample
of 253~fb$^{-1}$.
The strategy is 
to first identify a highly energetic lepton (electron or muon) and then
check consistency of all the other particles in the event with the hypothesis
that they come from the decay of a $B$ meson, by defining an acceptance
window in the $\Mbc$--$\DeltaE$ plane.
Continuum background is reduced by 
imposing requirements on the transverse component and the polar angle of the
missing momentum, corresponding to the neutrino in the signal decay. 
Further suppression is obtained by exploiting the event shape difference 
between continuum and $\BBbar$ events.
The main variable used to select signal is the lepton momentum in 
the $\bsig$ rest frame, $p_\ell^B$, which for the signal is expected to be
approximately equal to half of the $B$ meson mass.
A plot of $p_\ell^B$ for the muon mode after all selections are applied is
shown in Fig.~\ref{fig:belle-b2lnu} (left). Similar distributions are found
for the electron mode.
The signal yield is extracted from a fit to the $\btag$ $\Mbc$ distribution.
The distribution and fit curves are shown in 
Fig.~\ref{fig:belle-b2lnu} (right) for the muon mode. No evidence of signal 
is found in any of the modes, and the following upper limits are 
obtained on the branching fractions:
\begin{eqnarray}
 \mathrm{BF}(B^+\to\mu^+\nu_\mu) &<& 1.7 \times 10^{-6} ~~(90\,\%~\mathrm{CL}) \\
 \mathrm{BF}(B^+\to e^+ \nu_e) &<& 9.8 \times 10^{-7} ~~(90\,\%~\mathrm{CL}),
\end{eqnarray}
including the effect of the systematic uncertainties. 

\begin{figure*}[t!]
\centering
\includegraphics[width=80mm]{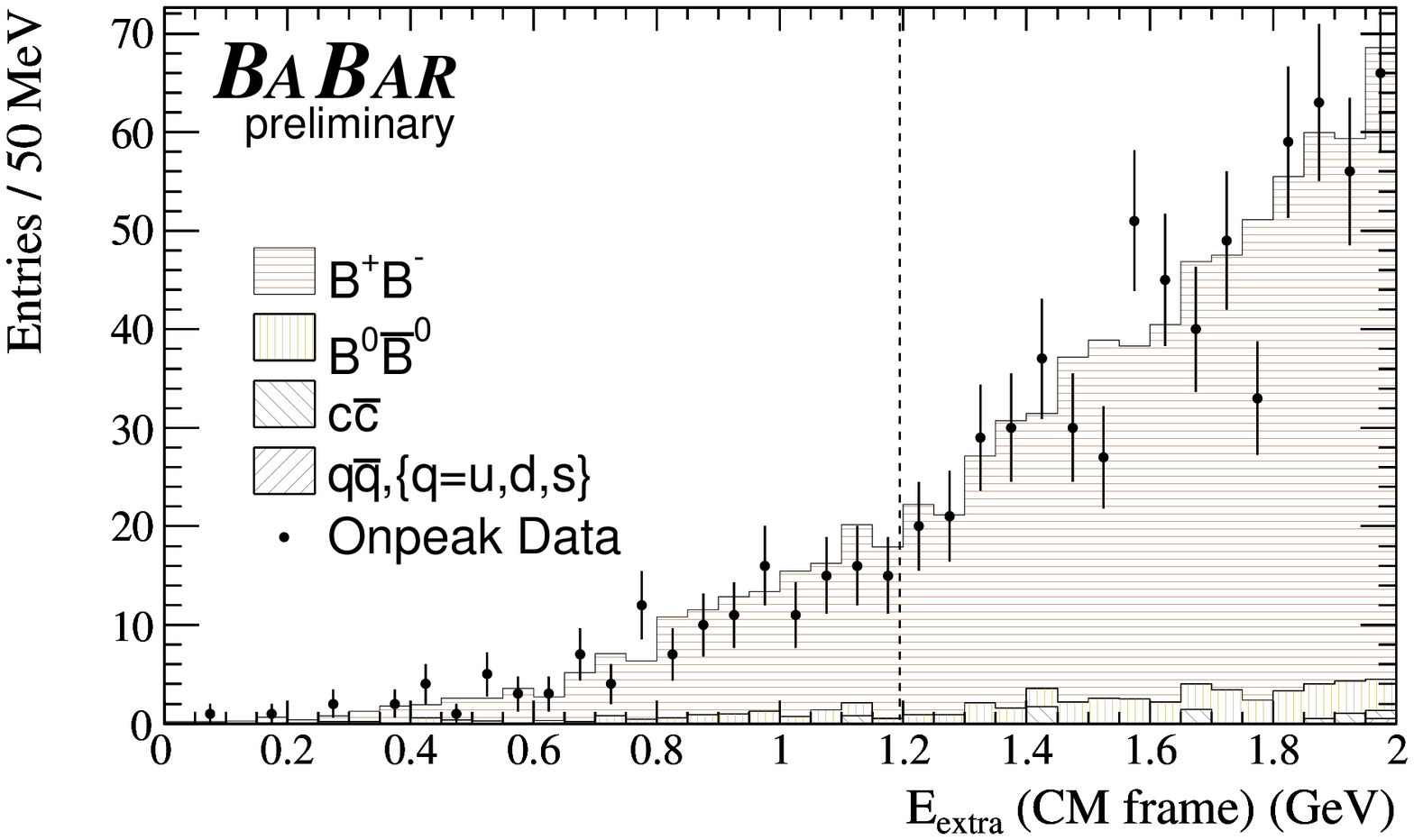}
\includegraphics[width=80mm]{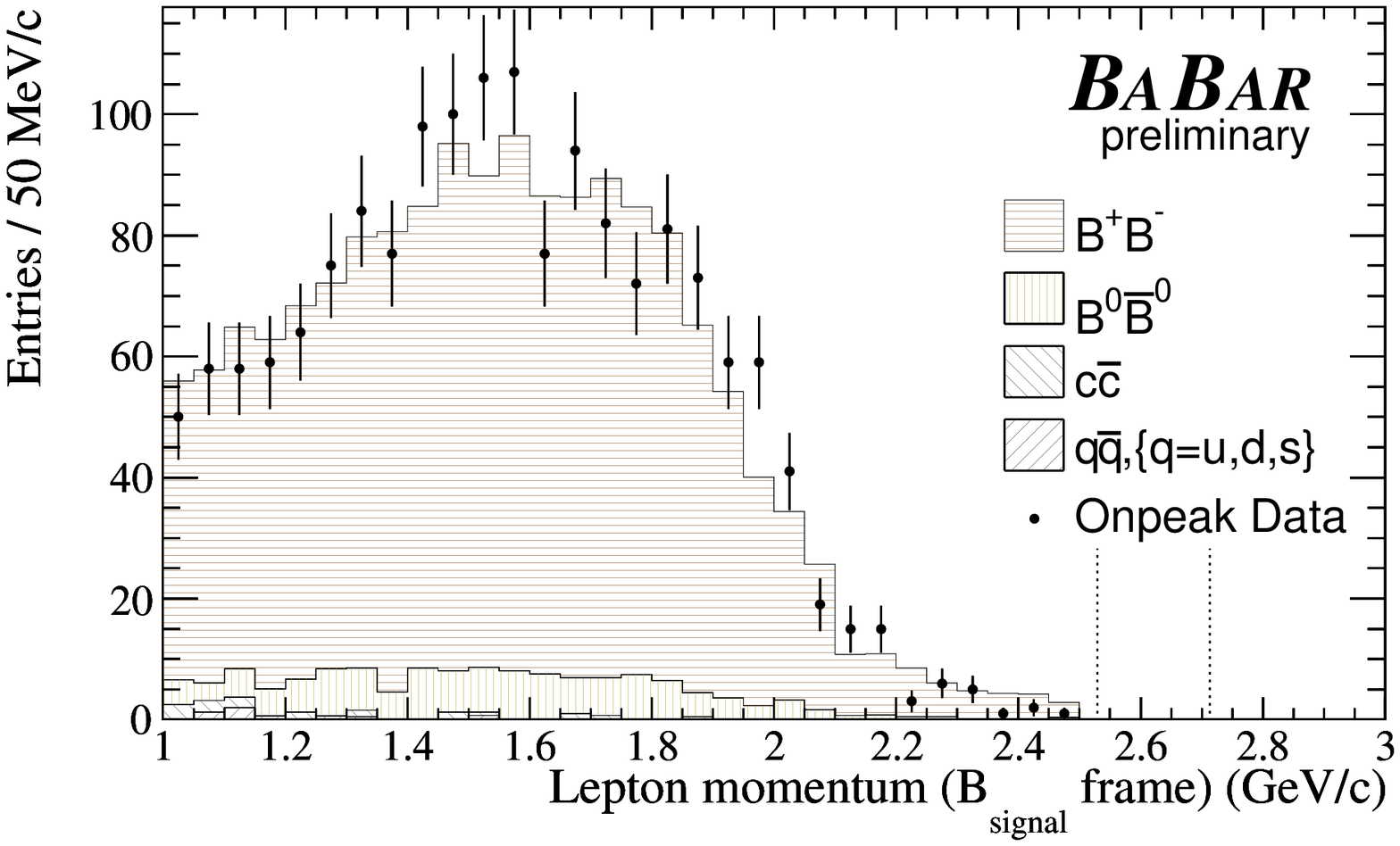}
\caption{Distribution of $E_\mathrm{extra}$ (left) and 
of the lepton momentum in the $B_{sig}$ rest frame (right) for the $\blnu$
analysis in BaBar. Histograms show the different components of the 
expected backgrounds, points are the selected events in the data.} 
\label{fig:babar-b2lnu}
\end{figure*}

The Babar's analysis~\cite{babar-b2lnu} is based on a sample of 209~fb$^{-1}$, and proceeds 
through a full reconstruction of the $\btag$ in the modes 
$B^- \to D^{(\ast)0} X^-$, where $X$ can contain any number of pions and
kaons. The signal selection and background suppression are carried out 
in a similar fashion to what described for the Belle's analysis above.
Events with $\eextra < 1.2 \GeV$ are finally selected 
(see Fig.~\ref{fig:babar-b2lnu}(left)) and a signal window is defined in the 
distribution of signal lepton momentum in the $\bsig$ rest frame.
No data points are selected in the signal window 
(see Fig.~\ref{fig:babar-b2lnu}(right)) and this leads to the following
upper limits, including systematic uncertainties:
\begin{eqnarray}
 \mathrm{BF}(B^+\to\mu^+\nu_\mu) &<& 6.2 \times 10^{-6} ~~(90\,\%~\mathrm{CL}) \\
 \mathrm{BF}(B^+\to e^+ \nu_e) &<& 7.9 \times 10^{-6} ~~(90\,\%~\mathrm{CL}).
\end{eqnarray}
%

%
%%%%%%%%%%%%%%%%%%%%%%%%%%%%%%%%%%%%%%%%%%%%%%%%%%%%%%%%%%%%%%%%%%%%%%%%%%%%%%%
\section{\boldmath Searches for $\blnug$}
\begin{figure}[ht!]
\centering
\includegraphics[width=80mm]{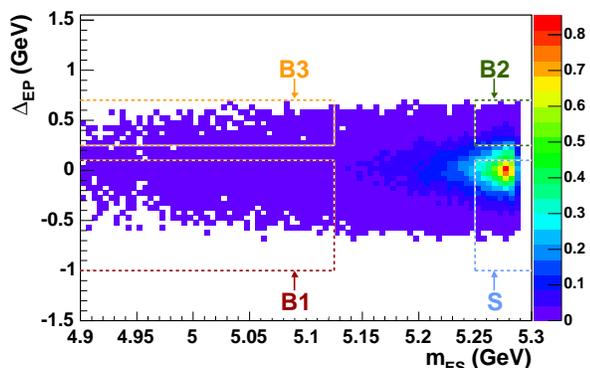}
\caption{Search for $\blnug$ in BaBar. Electron channel $\Delta_{EP}$ versus $m_\mathrm{ES}$ distribution
from signal MC. The signal (S) and the three sideband 
regions (B1, B2, B3) are drawn as dashed rectangles.
The color scale represents the relative content of each bin.} 
\label{fig:babar-b2lnugamma}
\end{figure}

%%%%%%%%%%%%%%%%%%%%%%%%%%%%%%%%%%%%%%%%%%%%%%%%%%%%%%%%%%%%%%%%%%%%%%%%%%%%%%%
%
The presence of the photon in decays of the type $\blnug$ can lift the helicity 
suppression that has so far prevented the observation of the $\blnu$ decay
in electronic and muonic modes. The $\blnug$ BF's are therefore independent
of lepton flavour up to factors of the order $(m_\ell / m_B)^2$.
The disadvantage of this type of decays is that the theoretical description
of the radiative process is not as straightforward as the one outlined in 
Sec.~\ref{Sec:blnu}, and therefore interpretation of results cannot be
obtained without some degree of model dependence. 
Searches for $\blnug$  decays with $\ell$ a muon or an electron, have been
performed by CLEO~\cite{cleo-blnug}, using a model for the signal~\cite{burdman} 
that predicted BF in the range 1--$4 \times 10^{-6}$.
Examining a data sample corresponding to 2.5~fb$^{-1}$, they extract the 
following upper limits at 90\% CL:
\begin{eqnarray}
 \mathrm{BF}(B^+\to\mu^+\nu_\mu \gamma) &<& 5.2 \times 10^{-5}  \\
 \mathrm{BF}(B^+\to e^+ \nu_e \gamma) &<& 2.0 \times 10^{-4} .
\end{eqnarray}
Preliminary results on the same channels have also been recently presented by 
BaBar, though based on different theoretical 
assumptions~\cite{babar-b2lnugamma, korchemsky}.
BaBar measures a partial BF ($\Delta$BF) 
in a restricted region of the phase space defined by
$1.875 < E_\ell < 2.850 \GeV$, $0.45 < E_\gamma < 2.35 \GeV$ and 
$\cos \theta_{\ell \gamma} < -0.36$, where $E_\ell$ and $E_\gamma$ are the 
CM-energies of the signal lepton and photon 
and $\theta_{\ell \gamma}$ is the angle between them evaluated in the CM frame.
In a sample of 210.5~fb$^{-1}$, the signal is formed selecting the highest 
CM-energy lepton and the highest CM-energy photon candidates; remaining particles
in the event are used for an inclusive reconstruction and selection of the recoiling
$\btag$.
After background rejection, the signal is extracted by maximizing a likelihood
function defined by counting data events in four regions (one signal
region and three sidebands, as shown in Fig.~\ref{fig:babar-b2lnugamma}) 
in the plane formed by the $\btag$ $m_\mathrm{ES}$ and by the difference between
the reconstructed neutrino candidate's energy and the magnitude of its 
3-momentum in the CM frame ($\Delta_{EP}$).
No excess of events is observed 
over the expected background, and the following 90\% CL Bayesian 
upper limits are extracted for $\Delta$BF, assuming flat priors in BF, 
for separate channels and for their combination:
\begin{eqnarray}
 \Delta\mathrm{BF}(B^+\to\mu^+\nu_\mu \gamma) &<& 2.1 \times 10^{-5}  \\
 \Delta\mathrm{BF}(B^+\to e^+ \nu_e \gamma) &<& 2.8 \times 10^{-5}    \\
 \Delta\mathrm{BF}(B^+\to \ell^+ \nu_\ell \gamma) &<& 2.3 \times 10^{-5} .
\end{eqnarray}
Within the theoretical model of Ref.~\cite{korchemsky}, the last limit
can be translated to an upper limit on the total BF:
BF$(B^+\to \ell^+ \nu_\ell \gamma) < 5.0 \times 10^{-6}$ at 90\% CL.

\bigskip % extra skip inserted

\end{document}